\documentclass[12pt,a4paper, amssymb, amsmath]{article}
\usepackage{achemso}
\usepackage{epsfig}
\usepackage{graphicx}
\usepackage{indentfirst}
\usepackage{multirow}
\usepackage{latexsym}
\newcommand{\degree}{^\circ}
\title{
Strain-dependent modulation of conductivity in single layer transition-metal dichalcogenides
}

\author{M.~Ghorbani-Asl, S.~Borini, A.~Kuc and T.~Heine 
\\
School of Engineering and Science, Jacobs University Bremen, \\
Campus Ring 1, 28759 Bremen, Germany}

\date{\today}
\begin{document}
\maketitle
\begin{abstract}
Quantum conductance calculations on the mechanically deformed monolayers of MoS$_2$ and WS$_2$ were performed using the non-equlibrium Green's functions method combined with the Landauer-B\"{u}ttiker approach for ballistic transport together with the density-functional based tight binding (DFTB) method.
Tensile strain and compression causes significant changes in the electronic structure of TMD single layers and eventually the transition semiconductor--metal occurs for elongations as large as ~11\% for the 2D-isotropic deformations in the hexagonal structure.
This transition enhances the electron transport in otherwise semiconducting materials.
\end{abstract}

\section{Introduction}
The semiconductor industry rapidly approaches the performance limit of current silicon-based technology. Short-channel effects and gate leakage increase levels of heat dissipation as transistors are miniaturized. Silicon, being a 3D material, suffers stability issues and quantum size effects if layers approach a critical thinness of a few \AA. Possible long-term solutions to this problem involve replacing silicon with another material that would perform better at smaller scales.
2D materials are very attractive alternatives, and the recent progress in the field of graphene has demonstrated that nanotechnology is currently in the position to process 2D materials. The move to 2D materials also enables new functionalities, such as surface-coated devices or flexible electronics.

\begin{figure}
\begin{center}
\includegraphics[scale=0.22,clip]{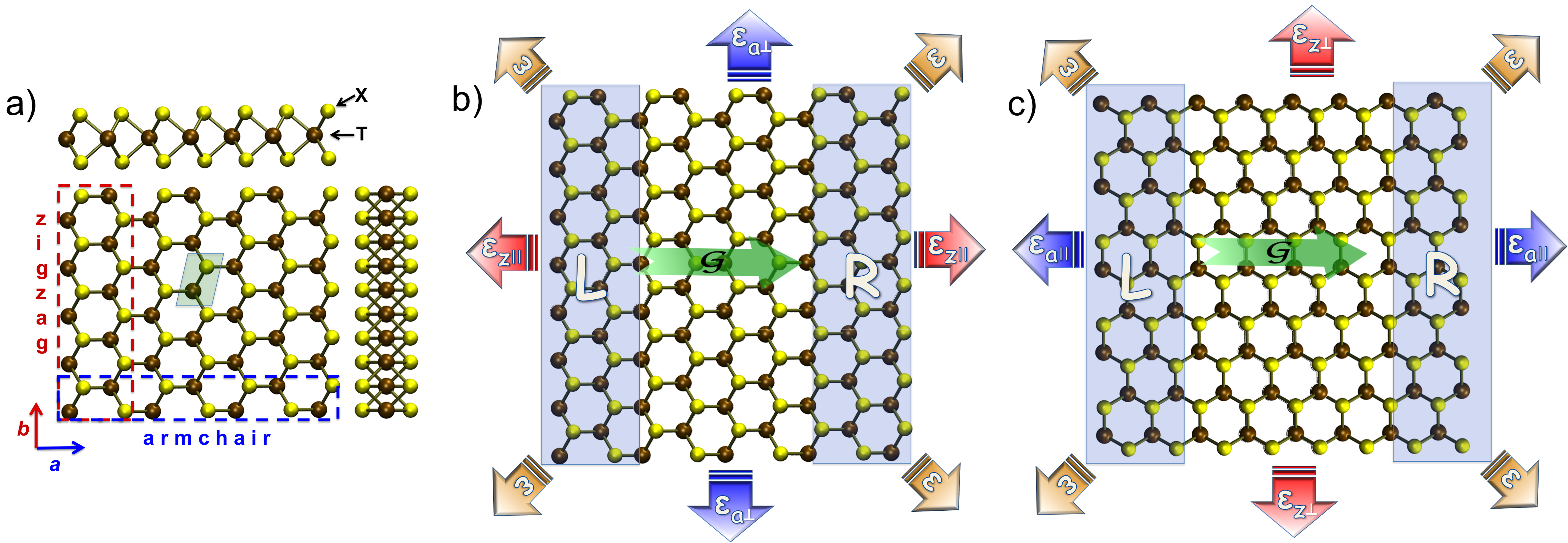}
\caption{\label{fig:1}(Online color) The atomic structure (a) of 2D transition-metal dichalcogenides of TX$_2$ type in the rectangular representation. The dashed rectangles indicate the units in $a$ and $b$ directions. The hexagonal unit cell is highlighted. Schematic representation of the electronic transport in the (b) zigzag and (c) armchair directions shown together with the possible mechanical deformations: 2D-isotropic ($\varepsilon$), uni-axial and parallel to the transport axis ($\varepsilon_{\parallel}$), uni-axial and perpendicular to the transport axis ($\varepsilon_{\perp}$). The electrodes, $L$ and $R$, are highlighted with shaded rectangles).}
\end{center}
\end{figure}

The drawback of graphene as a silicon alternative is the absence of a band gap and the difficulty to introduce one. There is, however, a rich variety of semiconducting 2D materials, most prominent are transition-metal dichalcogenides (TMDs) of TX$_2$ type (where T -- transition-metal atom, X -- chalcogen atom). They are composed of two-dimensional X-T-X sheets stacked on top of one another, as shown in Fig.~\ref{fig:1}.
Each T atom is covalently bonded to six X atoms located in the top and bottom layers forming a sandwiched material. 
In the bulk, adjacent slabs are held together by weak interlayer bonds, what leads to a quasi-2D character of TX$_2$. Exfoliation of the bulk down to the monolayer, however, changes the electronic properties of the materials by quantum confinement.\cite{Splendiani2010, Kuc2011, Komsa2012}
TMD monolayers possess high mobility of charge carriers, high thermal stability, and the presence of a distinct band gap. These intrinsic properties make them attractive for electronic applications.

Recently, Kis and coworkers demonstrated that these materials are indeed suitable for producing nanoelectrodevices, and the first field-effect transistors, logical circuits and amplifiers have been manufactured.\cite{Radisavljevic2011, Ghatak2011, Radisavljevic2011a, Radisavljevic2012} Large-scale production appears to be feasible, as liquid exfoliation of TMDs is able to produce large area two-dimensional monolayers.\cite{Coleman2011} 

Bulk layered TMDs can be either semiconductors (e.g. T=Mo, W) or metals (e.g. T=Nb, Re), depending on their stoichiometry. It has recently been shown that the former exfoliated TMDs in the form of monolayers undergo a transition of their electronic structure.\cite{Splendiani2010, Kuc2011}
While bulk structures are indirect band gap semiconductors, the 2D monolayers are direct band gap materials with significantly larger energy gaps of about 2~eV.\cite{Kuc2011, Mak2010}
The electronic structure of these TMDs can be further tuned by means of nanotube formation,\cite{Zibouche2012} substitutional doping\cite{Ivanovskaya2006, Ivanovskaya2008} or mechanical deformations.\cite{Johari2012, Yue2012, Yun2012, Scalise2012, Lu2012} The latter one is particularly interesting as the semiconductor-metal transition occurs at elongations of about 11\% without bond breaking. Such rather large elongations have been observed in the experiment, e.g. in tensile tests of WS$_2$ nanotubes.\cite{Kaplan-Ashiri2006}

The proposed semiconductor-metal transition\cite{Johari2012, Yue2012, Yun2012, Scalise2012, Lu2012} has been discussed so far solely by band structure calculations. However, stretched semiconductors are likely to expose states at the Fermi level due to broken (dangling) bonds, or bonds that are about to break, that do not confirm electrical conductance. It is imperative to study the electron conductance through these materials in order to conclude if 2D TS$_2$ layers could find applications as nanoelectromechanical switches. Here, we analyze strain effects on the electronic transport properties of two-dimensional TMDs using density functional based methods and 2D quantum transport calculations.

\section{Methods}
\label{Sec:Methods}

The fully optimized structures of TS$_2$ with T = Mo, W, and Nb were taken from our previous work.\cite{Kuc2011}
For computational details and optimized lattice parameters of TMDs, the reader is referred to the work of Kuc et al.\cite{Kuc2011}. In short: all stability and band structure calculations have been carried out using density-functional theory (DFT) employing the PBE (Perdew-Burke-Ernzerhof) functional.\cite{PBE} The equilibrium structures were exposed to mechanical strain following three possible strategies: 1) 2D-isotropic deformation ($\varepsilon$) on the hexagonal unit cells with simultaneous change of $a$ and $b$ lattice vectors, 2) uni-axial deformation of $a$  along the armchair edges ($\varepsilon_{a}$) in rectangular unit cell, and 3) uni-axial deformation of $b$ along the zigzag edges ($\varepsilon_{b}$) in the rectangular unit cell. Depending on the direction of the current, the uni-axial deformations can be either perpendicular ($\varepsilon_{\perp}$) or parallel ($\varepsilon_{\parallel}$). For example, the notation of $\varepsilon_{a\perp}$ denotes the uni-axial strain along $a$ lattice vector and perpendicular to $\mathcal{G}$ (see Fig.~\ref{fig:1}). Note, that the uni-axial tensile strain along the $b$ and $a$ vectors correspond also to the deformations of large-diameter zigzag and armchair TMD nanotubes, respectively. The strain is defined as $\varepsilon=\frac{L-L_{0}}{L_{0}}$, where ${L_{0}}$ and ${L}$ are equilibrium and strained lattice values.

The coherent electronic transport calculations were carried out using our in-house non-equilibrium Green's functions (NEGF)\cite{Datta2005} code combined with the Landauer-B\"{u}ttiker approach, that takes advantage of the density functional based tight-binding (DFTB) Hamiltonians.\cite{BAND-DFTB, Oliveira2009, Seifert1996} The Slater-Koster parameters for DFTB calculations were created following the approach described by Oliveiara et al.~\cite{Oliveira2009} and Seifert et al.,\cite{Seifert1996} and have been validated by band structure comparison with DFT calculations. [For the band structures calculated at the DFT/PBE level see Supporting Information].\cite{SI} 
We have used the PBE\cite{PBE} density-functional, numerical basis functions, and the quadratic confinement potential with $r_0$ = 5.5, 5.8, 5.9, and 3.8 Bohr for Mo, W, Nb, and S, respectively. DFTB band structure were calculated using DFTB+ code.\cite{dftb+} The mean free path of the carriers in TMDs is greater than the sample dimensions and their high mobility\cite{Radisavljevic2011} implies that the carriers can move long distances without scattering. Hence, it is expected that the ballistic approximation is appropriate for studying transport in TMDs.\cite{Yoon2011}

The electron conductivity was calculated using a system that consists of a finite TMD monolayer as scattering region (conductor), which is connected to two semi-infinite TMD monolayers that act as electrodes, $R$ and $L$ (see Fig.~\ref{fig:1}). In this way, the so-called bulk conductivity is simulated, where the interface between the electrodes and the conductor is well defined. In the bulk conductivity, the scattering region has to be large enough to avoid the interaction between $R$ and $L$. We have converged the scattering region to seven unit cells between the electrodes. The direction perpendicular to the transport axis is assumed to be two dimensional by accounting that each border atom in the unit cell is surrounded by translational equivalence of atoms according to the symmetry of the periodic system. In this direction, the periodicity is described by the $\Gamma$-point approximation, recently justified for analogues systems.\cite{ghorbani2012} To avoid spurious overlap between atoms from the image cells, we have used twelve unit cells in the direction perpendicular to the current. In this way, the edge effects arising from the typical one-dimensional representation of the layers as nanoribbons are discarded from the simulations.
At low temperature, the electron conductivity through such a system can be described using Fisher-Lee Formula:\cite{Fisher1981}
\begin{equation}
\mathcal{G}(E)=\frac{2e^{2}}{h}\sum_n T_{n}(E)=\frac{2e^{2}}{h}Tr\left[\hat{G}^{\dagger}\hat{\,\Gamma}_{R}\hat{\, G}\hat{\,\Gamma}_{L}\right],
\end{equation}
where $\hat{G}$ is the Green's function of the whole system, and $\vphantom{}$$\hat{\Gamma}_{L,R}=-2\mathrm{Im\mathit({\hat{\Sigma}_{L,R})}}$ indicate the coupling matrices.
Self-energies, ${{\hat\Sigma}_{L,R}}$, which describe the influence of the semi-infinite electrodes on the electronic structure of the device, can be calculated via a self-consistent procedure.\cite{Lopez1985}

Hirshfeld charge calculations were performed using the ADF-BAND code, with PBE functional and triple-zeta basis sets with one polarization function (TZP), including scalar ZORA relativistic correction.\cite{BAND}

The elastic properties of the monolayers under the uni-axial deformations correspond to those of TMD nanotubes with large diameters.
Therefore, we have used the tensile stress as force $F$ acting on the area $A$. For the area calculation, we have used the fixed lattice vector (e.g.\ if $a$ was deformed, $b$ was used to calculate A), multiplied with the optimized monolayer thickness of 6.2\AA.

\section{Results and Discussion}
\begin{figure}
\begin{center}
\includegraphics[scale=0.45,clip]{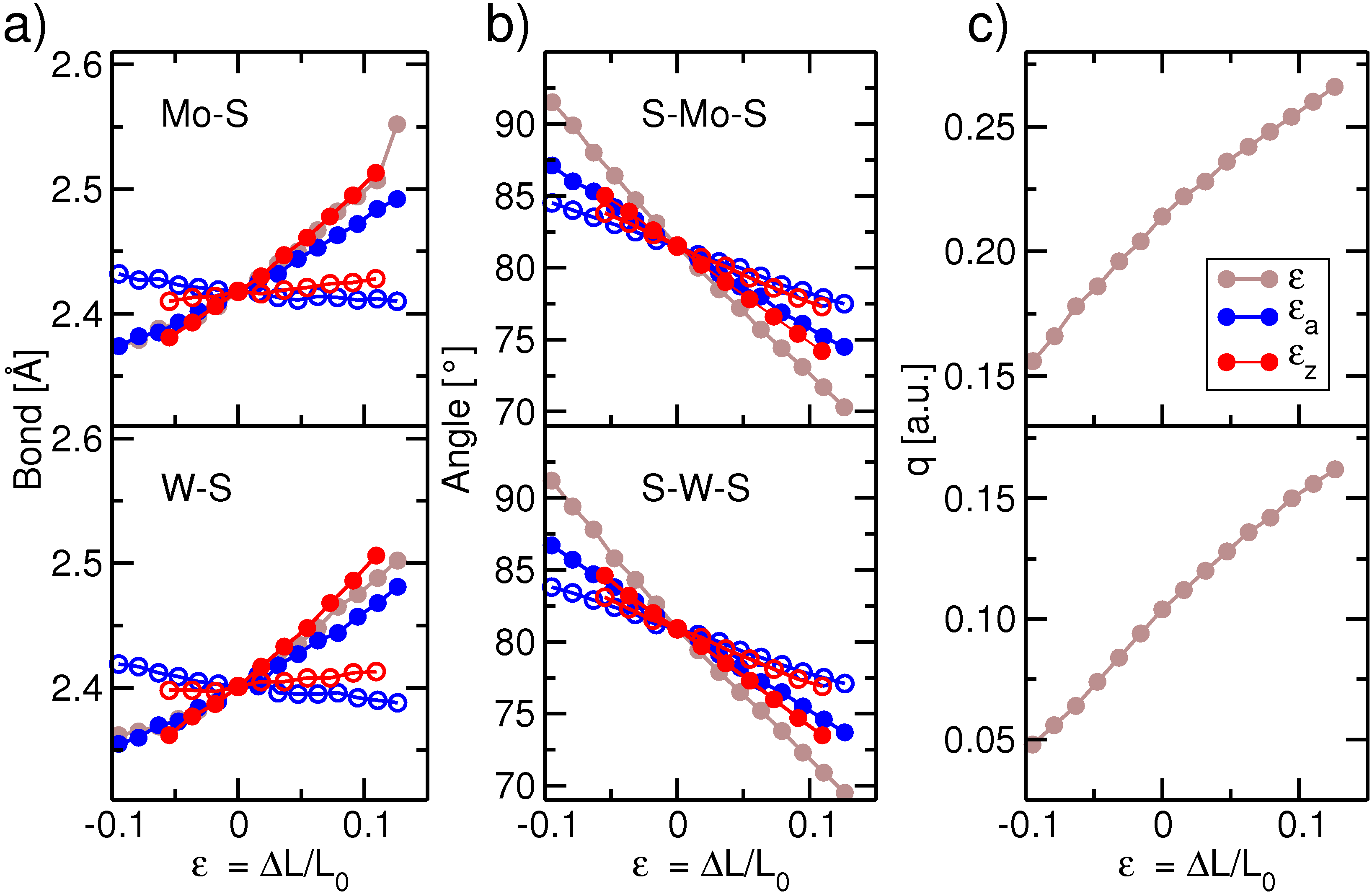}
\caption{\label{fig:2}(Online color) Calculated T--S bond lengths (a), S--T--S bond angles (b), and the T atomic charges (c) of MoS$_2$ (top) and WS$_2$ (bottom) monolayers under tensile strain and compression. Full symbols denote the variables along the deformation axis, empty symbols correspond to the variables along the fixed axis.}
\end{center}
\end{figure}

Tensile stress and compression within the monolayers cause changes in the geometry, namely the elongation or shrinking of the bond lengths and bond angles.
TMDs can be exposed to such strong mechanical deformations without breaking the T-S bonds due to very flexible S--T--S bond angles, which can be easily  stretched or compressed by up to about 10$\degree$. 
Figs.~\ref{fig:2} a and b show the calculated interatomic distances (T--S) and bond angles (S--T--S) as function of the applied strain.
T--S bond distances become longer by about 0.01\AA\ per 1.5\% isotropic tensile strain.
At the same time, significant changes are obtained in the S--T--S bond angles: about 1.5$\degree$ per 1.5\% isotropic tensile strain.
Both, the bond lengths and angles scale nearly linearly with the applied deformation for up to about 11\% strain/compression.
Uni-axial deformation cause slower changes in the geometry for a given $\varepsilon$ comparing with the isotropic deformations.

Changes in the geometry of TMD monolayers under mechanical deformations also affect the electronic structure of these materials.
For isotropic deformations, we have calculated evolution of the atomic charge (q) for the T atoms using Hirshfeld atomic charge analysis.
The Mo and W atomic charges are shown in Fig.~\ref{fig:2} c as function of $\varepsilon$.
The transition metal atoms become more positive with increasing the tensile strain, at the same time the sulfur atoms get more negative.
This leads to a charge depletion around the T atoms and consequently charge excess around S atoms.
Atomic charge analysis also let us discuss the charge transfer (CT) from the T to S atoms, which also increases with the applied strain.
Therefore, the T--S bonds become more ionic in comparison with the equilibrium structures.
In addition, the differences in the atomic charges of both metals suggest that MoS$_2$ is more polarized compared with WS$_2$.

\begin{figure}
\begin{center}
\includegraphics[scale=0.30,clip]{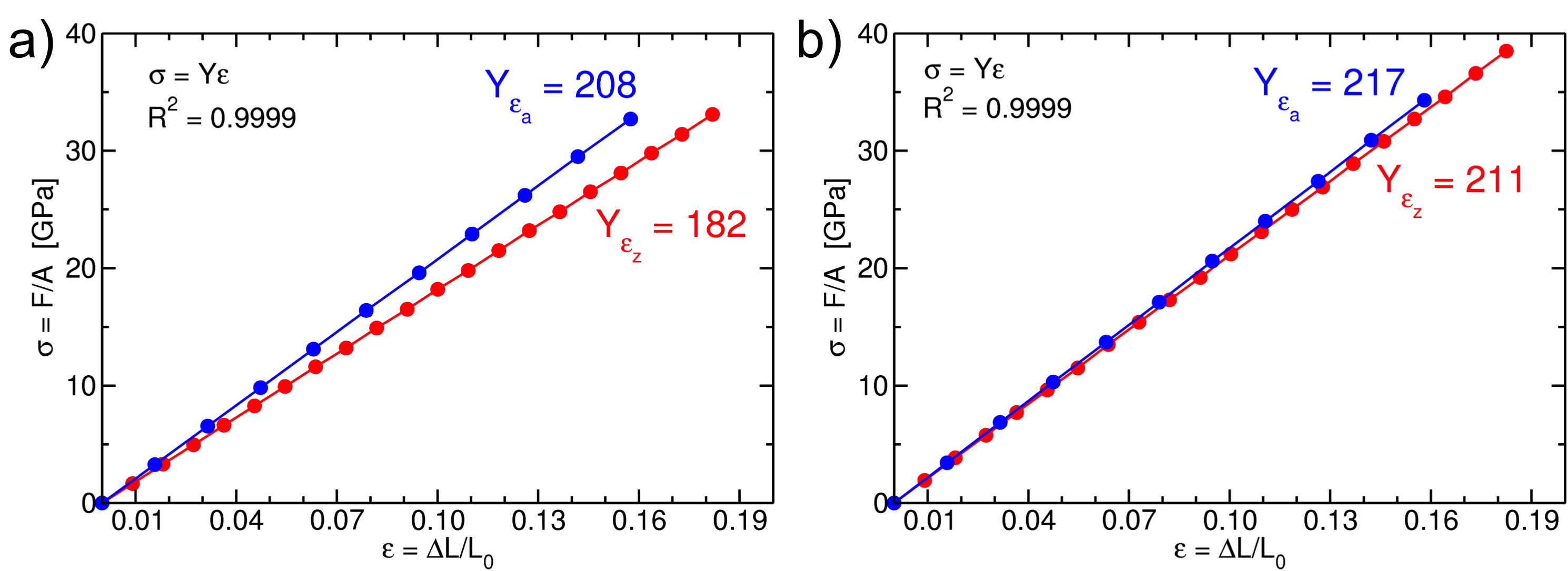}
\caption{\label{fig:3}(Online color) Calculated  strain-stress relation of MoS$_2$ (a) and WS$_2$ (b) monolayers under uni-axial tensile strain. The Young's moduli for the deformations along the $a$ and $b$ lattice vectors are indicated by $Y_{\varepsilon_{a}}$ and $Y_{\varepsilon_{z}}$ for the armchair and zigzag edge types, respectively.}
\end{center}
\end{figure}

The geometry changes due to the applied deformations are elastic and the materials should return to their equilibrium structures once the strain is released.
Compliance with Hooke's law is confirmed by our calculated stress-strain relation curves, which show linear scaling within the range of applied $\varepsilon$.
The stress-strain curves for the uni-axial deformations are shown in Fig.~\ref{fig:3}.
The observed linear relation means that we are working in the truly elastic region, and that the TMDs are very ductile materials.
Fitting the linear function to the stress-strain curve gives the corresponding elastic modulus. 
For the MoS$_2$ monolayer, this elastic modulus is 208 GPa and 182 GPa for the armchair and zigzag directions, respectively.
Our results are in very good agreement with the recent work of Li et al.,\cite{Tianshu2012} who have reported  200.3$\pm$3.7 GPa and 197.9$\pm$4.3 GPa, respectively, using DFT/PW91 level of theory as implemented in PWscf code.
We note that our elastic modulus is 22 GPa smaller than the Young's modulus reported by Lorenz et al.\cite{Lorenz2012} for the MoS$_2$ nanotubes calculated at the DFTB level.  
This deviation can be related to the differences in the optimized monolayer thickness and the quantum-mechanical method used for the computations.
The stress-strain relation and the elastic modulus obtained for WS$_2$ is nearly the same for the zigzag and armchair directions (217 vs. 211 GPa). 

\begin{figure}
\begin{center}
\includegraphics[scale=0.32,clip]{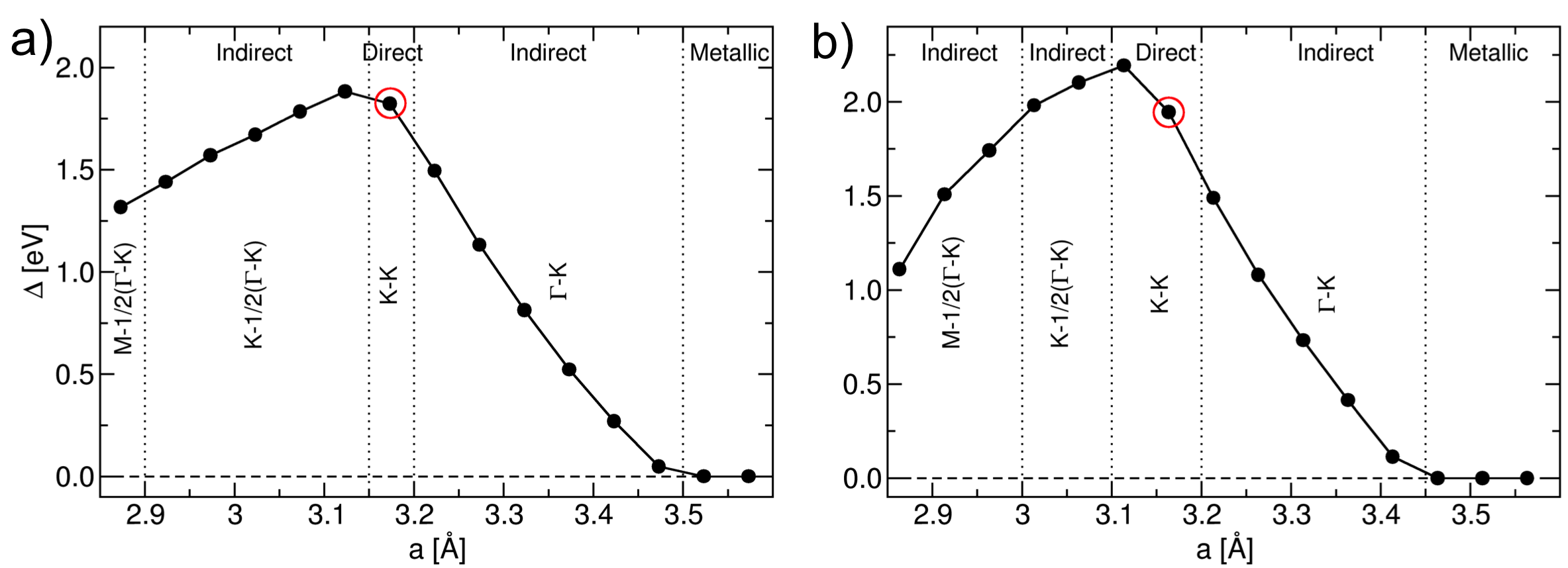}
\caption{\label{fig:4} Band gap evolution of MoS$_2$ (a) and WS$_2$ (b) monolayers under isotropic tensile strain and compression. The $k$-points, at which the top of valence band and the bottom of conduction band occur, are indicated. The equilibrium structures are marked with red circles.}
\end{center}
\end{figure}

\begin{figure}
\begin{center}
\includegraphics[scale=0.40,clip]{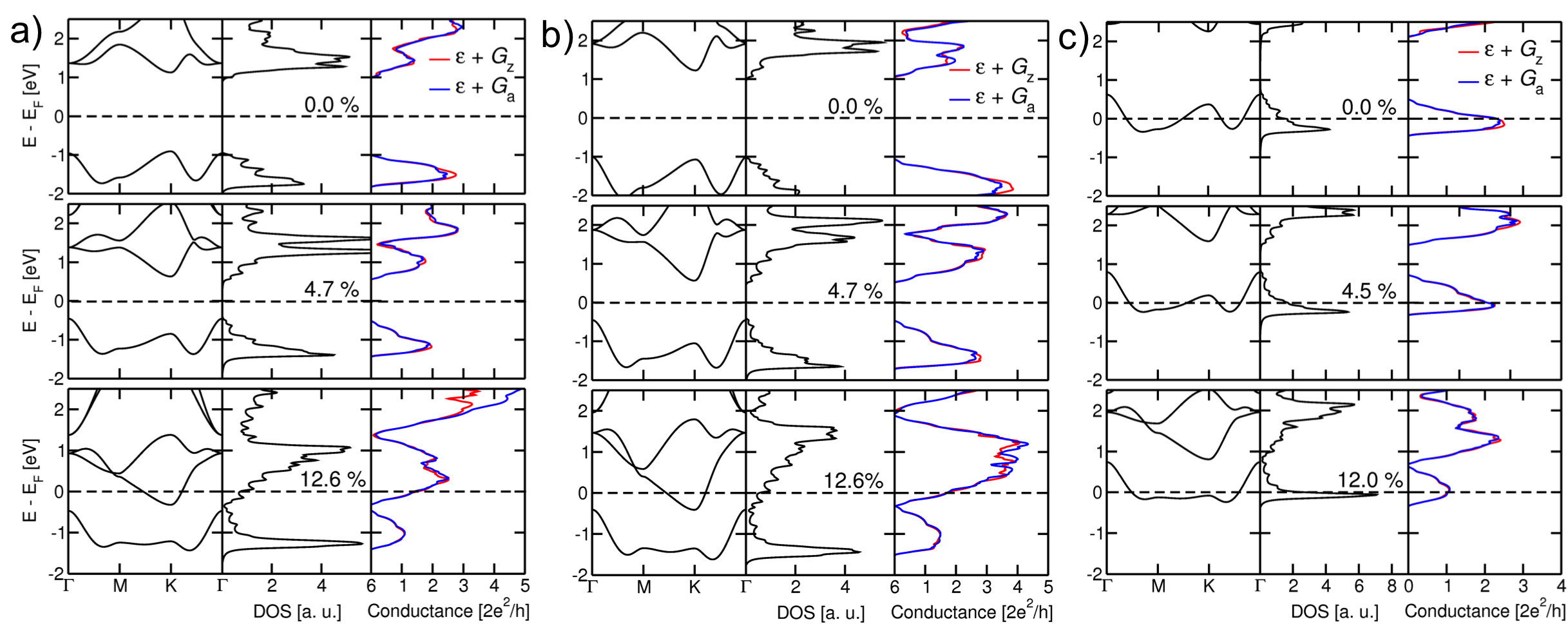}
\caption{\label{fig:5}(Online color) Band structure, the corresponding density of states, and the electron conductivity of MoS$_2$ (a) WS$_2$ (b), and NbS$_2$ (c) monolayers under isotropic tensile strain. $\mathcal{G}_a$ and $\mathcal{G}_z$ denote electron conductivity along the armchair and zigzag direction, respectively.}
\end{center}
\end{figure}

Fig.~\ref{fig:4} shows the band gap evolution with the applied 2D-isotropic strain for MoS$_2$ and WS$_2$ hexagonal monolayers. In equilibrium, the single TMD layers are direct band gap semiconductors with the transition at the $K$ point. Under tensile strain, the band gap decreases linearly and the system exhibits an indirect transition from the $\Gamma$ to the $K$ point. Eventually, the band gap vanishes for elongations of about 11\%, and consequently the monolayers become metallic with bands crossing the Fermi level at $K$ (conduction band) and $\Gamma$ (valence band) points. At this elongation the T--S bonds are stretched, but not yet broken (cf.\ Fig.~\ref{fig:2}). Similar behavior is observed when a monolayer is exposed to compressive strain, though the influence is smaller than that of the tensile strain. Uni-axial tensile strain results in a similar band gap lowering with eventual transition to the metallic state. In this case, however, for a given amount of elongation or compression, the changes are smaller than those observed for 2D-isotropic deformations. These results are in a very good agreement with the works of Johari et al.,\cite{Johari2012} Yue et al.,\cite{Yue2012} and Peelaers et al.\cite{Peelaers2012}
The influence of mechanical deformations on the electronic structure is mostly visible at the $K$ point, where both the bottom of conduction and the top of valence bands are significantly lowered compared with the equilibrium situation. At the same time, the top of valence band at the $\Gamma$ point moves upwards (see Fig.~\ref{fig:5}). This behavior can be understood as the orbital overlap between the Mo and S atoms decreases due to the change in the bond distances.\cite{Yue2012}

The changes in the electronic structures under mechanical deformations can enhance or quench the electron conductivity through semiconducting TMD monolayers. Fig.~\ref{fig:5} shows the comparison of electronic  structures and the corresponding electron conductivities of MoS$_2$ and WS$_2$ monolayers. Three examples are plotted, namely direct band gap in the equilibrium structure, indirect band gap (4.7\% elongation), and the metallicity (12.6\% elongation).
The 2D-isotropic strain ($\varepsilon$) gives rise to reduction of the band gap and significant increase in the electron conductivity at the Fermi level.

Since at approximately 11\% deformation band gaps of MoS$_2$ and WS$_2$ vanish, the transport channels are completely open and electrical conductivity is enabled. This can be explained by the deformation of the Brillouin zone, where the applied strain uniformly changes the primitive cell, bands cross the Fermi level, and consequently the character of a monolayer changes from the semiconducting to metallic.
The electronic transport in the case of isotropic strain/compression does not depend on the direction of the current and it is almost unchanged for the zigzag ($\mathcal{G}_z$) and the armchair ($\mathcal{G}_a$) directions.

\begin{figure}
\begin{center}
\includegraphics[scale=0.38,clip]{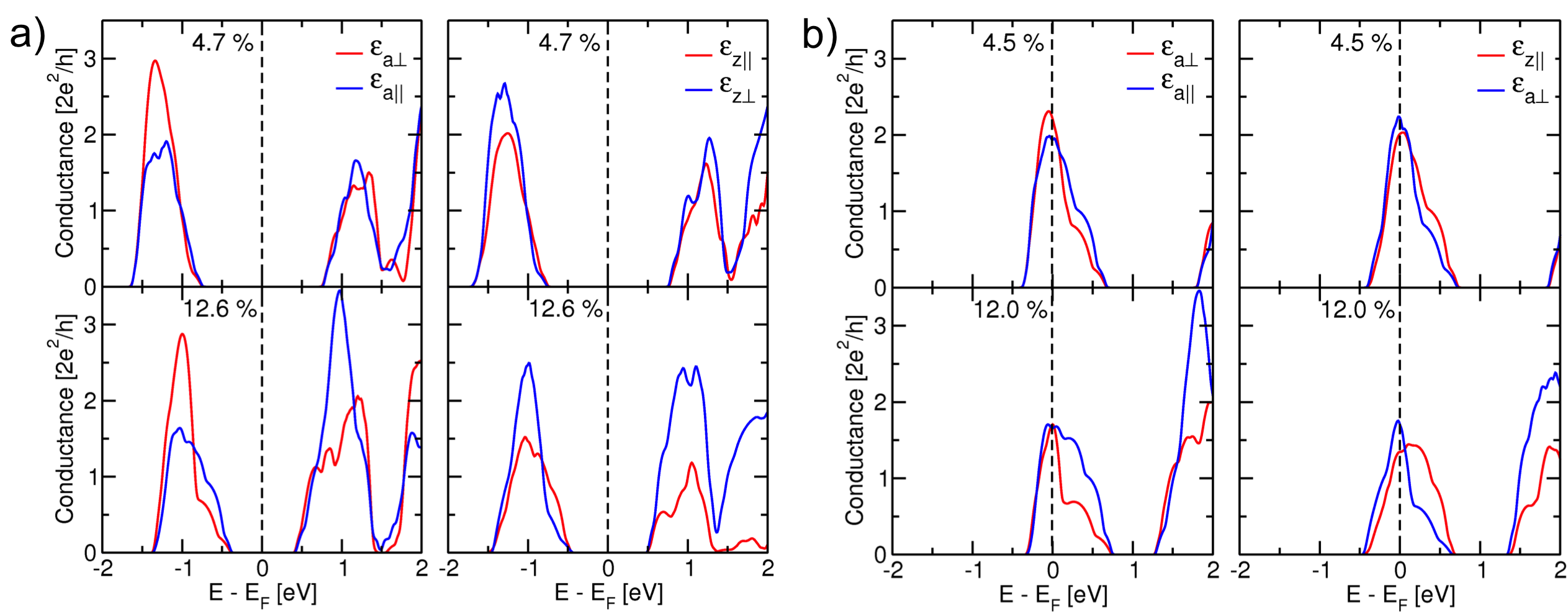}
\caption{\label{fig:6}(Online color) Electron conductivity of MoS$_2$ (a) and NbS$_2$ (b) monolayers under uni-axial tensile strain.}
\end{center}
\end{figure}

Fig.~\ref{fig:6} shows the electron conductivity evolution of MoS$_2$ monolayers with uni-axial strain applied in the parallel ($\varepsilon_{\parallel}$) and perpendicular ($\varepsilon_{\perp}$) directions to the transport axis.
For elongations as large as ~13\%, the monolayers still show semiconducting behavior.
This case corresponds to the stretching of TMD nanotubes with very large diameters and is consistent with our results on the mechanical deformation of these materials\cite{Ghorbani2013} and the recent results of Lu et al.\cite{Lu2012}
Nevertheless, $\varepsilon_{\perp}$ shows a very useful way of tuning the band gaps of TMD monolayers in a wide range without changing the conductivity significantly (see Fig.~\ref{fig:6} a).
In contrast, $\varepsilon_{\parallel}$ behaves differently and a decrease in the band gap result also in the reduced conductances for the electron energies lower than the Fermi energy.
At the energies higher than the Fermi level, $\varepsilon_{\parallel}$ enhances the conductance dramatically (see Fig.~\ref{fig:6} b).These results highlight the importance of the strain direction for nanoelectromechanical applications.

For comparison, we have also calculated the electronic transport of metallic NbS$_2$ monolayers under tensile strain. At the Fermi level, the electron conductivity is present independent of the mechanical deformation, however, the intensity decreases significantly for both isotropic and uni-axial tensile strain (cf. Figs.~\ref{fig:5} and \ref{fig:6}).

\section{Conclusion}

Using non-equilibrium Green's functions (NEGF)\cite{Datta2005} method combined with the Landauer-B\"{u}ttiker approach we have calculated quantum transport of the mechanically deformed MoS$_2$ and WS$_2$ monolayers.
Under 2D-isotropic strain, the TMD monolayers change the semiconducting direct gap character and become metallic for the deformations as large as 11\%, what results in electrical conductivity at the Fermi level.
Uni-axial tensile strain of monolayers has less influence on the electronic structure for a given elongation.
This corresponds also to the deformations of large-diameter TMD nanotubes with zigzag and armchair chiralities for the $\varepsilon_{z}$ and $\varepsilon_{a}$, respectively.
The semiconductor-conductor transition is reflected in the electron transport, our calculations show similar values for metallic NbS$_2$ as for stretched MoS$_2$ and WS$_2$.
We expect that our calculations for monolayers are directly transferable to TMD nanotubes, in particular for the experimentally available ones with rather large diameters.
A study on TMD nanotubes is in progress and will be a subject of a forthcoming communication.\cite{Ghorbani2013}

Generally, mechanical deformation is a possible way of tuning the band gaps and transport properties of TMD monolayers due to the metal-semiconductor transition.
Therefore, strain effects in TMDs can opens new prospects for their applications in optoelectronics or nanosensor device engineering.

\section{Acknowledgements}
This work was supported by the German Research Council (Deutsche Forschungsgemeinschaft, HE 3543/18-1), The European Commission (FP7-PEOPLE-2009-IAPP QUASINANO, GA 251149). We thank Prof. Gotthard Seifert (TU Dresden) for helpful discussions.

\providecommand*\mcitethebibliography{\thebibliography}
\csname @ifundefined\endcsname{endmcitethebibliography}
  {\let\endmcitethebibliography\endthebibliography}{}

\end{document}


\maketitle

\section*{Support Infromation}
\begin{figure}
\begin{center}
\includegraphics[scale=0.62,clip]{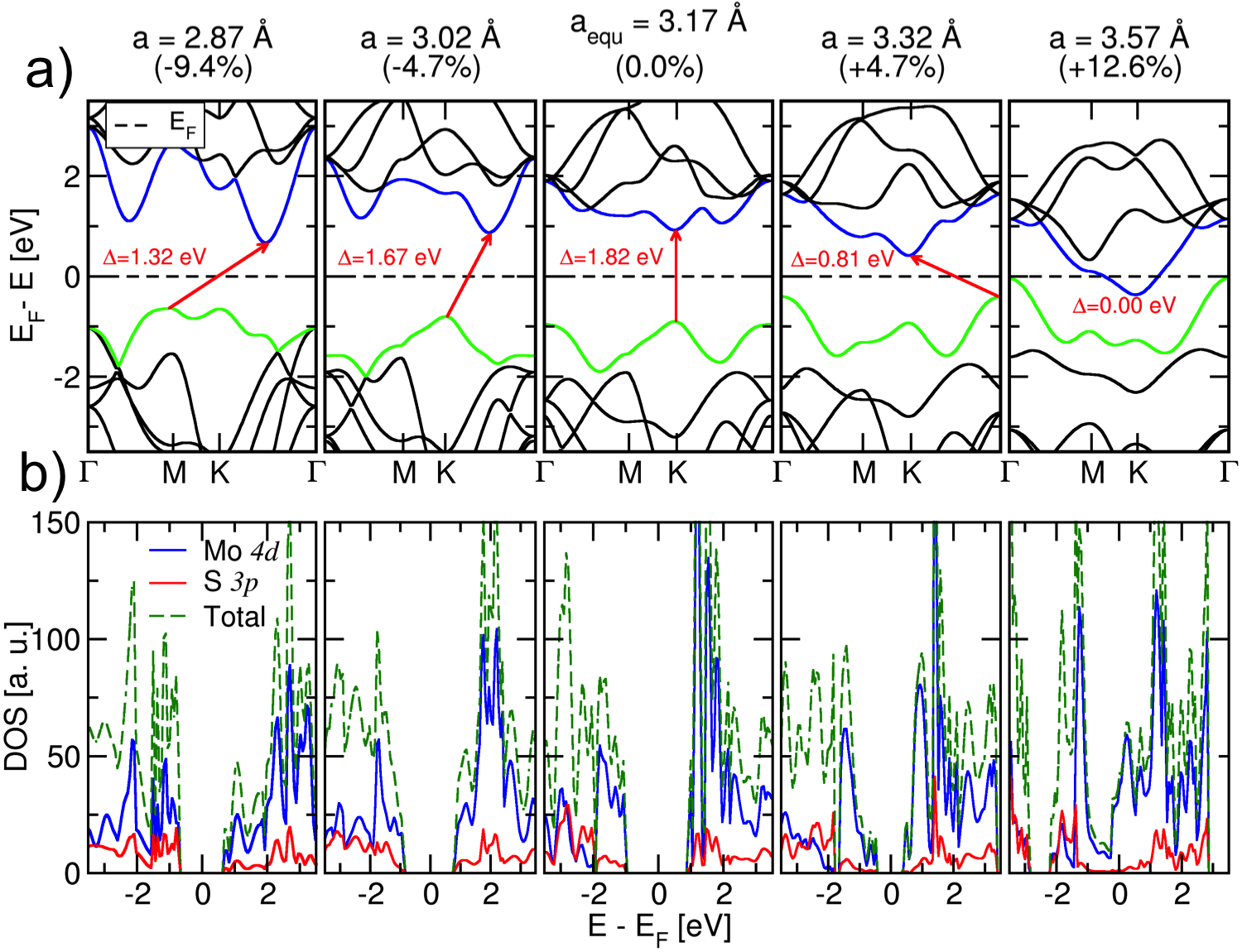}
\caption{\label{fig:1}(Online color) (a) Band structures of equilibrium and deformed MoS$_2$ monolayers calculated at the DFT/PBE level. The horizontal dashed lines indicate the Fermi level. The arrows indicate the fundamental band gap (direct or indirect) for a given system. The top of valence band (blue) and the bottom of conduction band (green) are highlighted. (b) Corresponding partial densities of states of MoS$_2$ monolayer with projections on the Mo and S atoms: the contributions from 4$d$ and 3$p$ orbitals of Mo and S, respectively.}
\end{center}
\end{figure}

\begin{figure}
\begin{center}
\includegraphics[scale=0.62,clip]{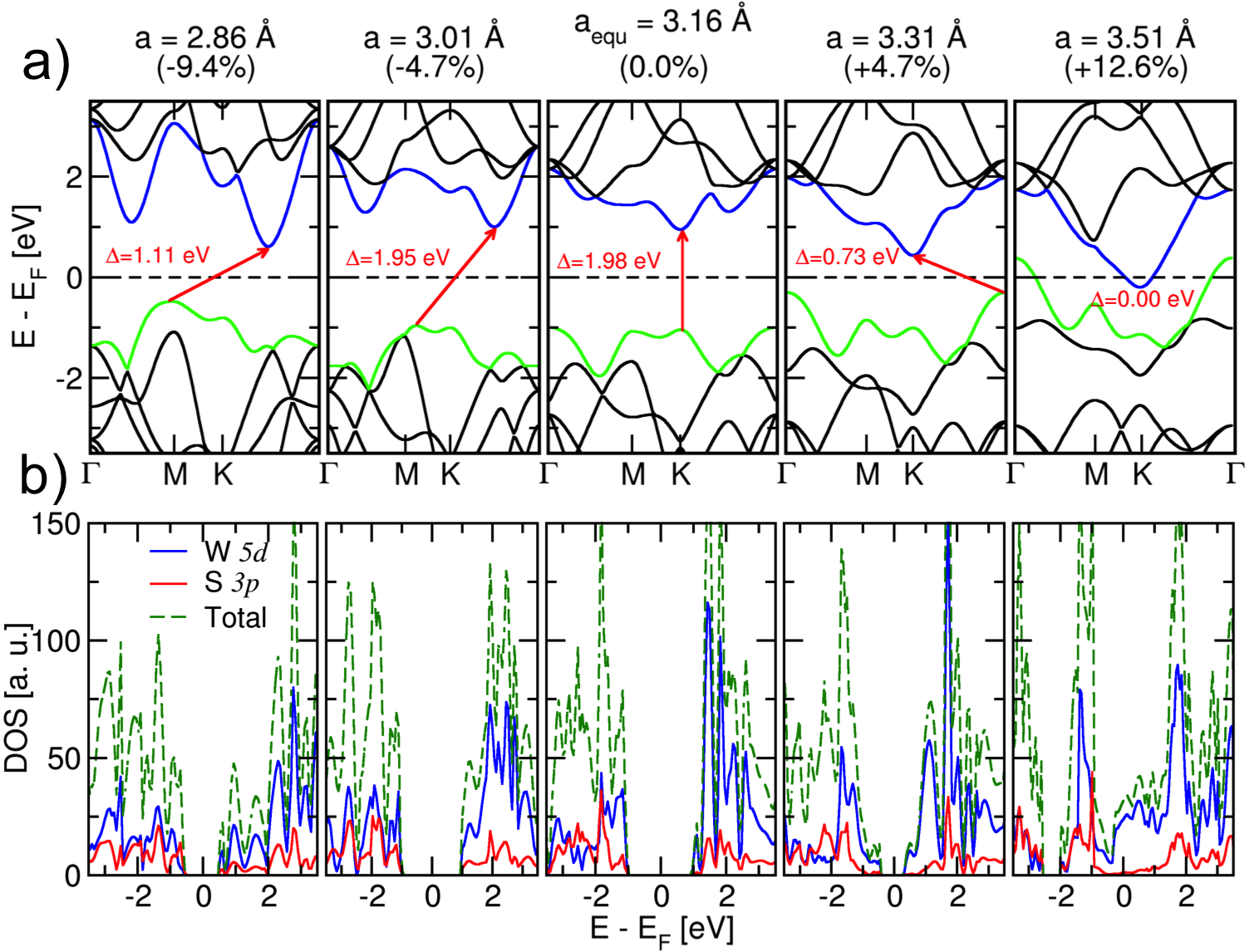}
\caption{\label{fig:2}(Online color) (a) Band structures of equilibrium and deformed WS$_2$ monolayers calculated at the DFT/PBE level. The horizontal dashed lines indicate the Fermi level. The arrows indicate the fundamental band gap (direct or indirect) for a given system. The top of valence band (blue) and the bottom of conduction band (green) are highlighted. (b) Corresponding partial densities of states of WS$_2$ monolayer with projections on the Mo and S atoms: the contributions from 5$d$ and 3$p$ orbitals of W and S, respectively.}
\end{center}
\end{figure}

Band structures and partial densities of states of TS$_2$  (T = Mo, W) monolayers under mechanical deformations calculated at  the DFT/PBE level (see Figs.~\ref{fig:1} and \ref{fig:2}).

